\documentclass[aps,pre,amsmath,lengthcheck,superscriptaddress]{revtex4-2}

\usepackage{graphicx}\graphicspath{ {figures/} }
\usepackage{hyperref}
\hypersetup{colorlinks,allcolors=blue,breaklinks}

\newcommand{\be}{\begin{equation}}
\newcommand{\ee}{\end{equation}}
\newcommand{\ba}{\begin{align}}
\newcommand{\ea}{\end{align}}
\newcommand{\bi}{\begin{itemize}}
\newcommand{\ei}{\end{itemize}}

\newcommand{\la}{\left\langle}
\newcommand{\ra}{\right\rangle}
\newcommand{\pd}{\partial}

\newcommand{\bla}{bla\\bla\\bla\\bla\\bla}

\newcommand{\mb}[1]{\mbox{\boldmath$#1$}}
\newcommand{\mc}[1]{\mathcal{#1}}

\begin{document}

\title{Global optimization and monotonicity in entropy production of weak drivings}

\author{Pierre Naz\'e}
\email{pnaze@unifap.br}
\affiliation{\it Departamento de Ci\^encias Exatas e Tecnol\'ogicas, Universidade Federal do Amap\'a, 68903-419, Macap\'a, Amap\'a, Brazil}

\date{\today}

\begin{abstract}

Knowing if an optimal solution is local or global has always been a hard question to answer in more sophisticated situations of optimization problems. In this work, for finite-time and weak isothermal driving processes, we show the existence of a global optimal protocol for the entropy production. We prove that by showing its convexity as a functional in the derivative of the protocol. This property also proves its monotonicity in such a context, which leads to the satisfaction of the Second Law of Thermodynamics. In the end, we exemplify that the analytical technique of the Euler-Lagrange equation applied to overdamped Brownian motion delivers the global optimal protocol, by comparing it with the results of the global optimization technique of genetic programming.

\end{abstract}

\maketitle

\section{Introduction}
\label{sec:introduction}

In real life, thermodynamic processes always occur in finite time \cite{deffner2020}, and energy is irreversibly spent in this manner \cite{callen1998}. An important question is naturally posed: is it possible to execute this same process with minimal dissipation? Mathematically this is nothing more than an optimization problem, where a physical quantity will be minimized in a control parameter under certain conditions. A gas contained in a box, where its volume is changed to accomplish some goal, is a typical example of this scenario.

Several tools have been developed to treat problems of optimization: from the basic techniques of calculus of variations and optimal control theory \cite{gelfand2000,kirk2004} to numerical techniques like algorithms in convex, global and nonlinear optimizations \cite{boyd2004,weise2009,ruszczynski2011}. In any of those scenarios, questions involving the global minimum, that is, the least minimum among all possible solutions, are always present. Even though answers are hard to find in general, a simple criterion always used to guarantee their existence is the idea of convexity. It affirms that if a function is convex, any local minimum will be always a global one \cite{boyd2004}. Similarly, the existence of a global maximum is guaranteed with the concept of concavity. 

Although those concepts seem like pure mathematical ideas, they have important physical consequences. In classical Thermodynamics, for instance, the concavity of the entropy with respect to the energy guarantees the existence of a unique new equilibrium state to where the system converges when the constraints are changed \cite{callen1998}. Also, since a convex function is always monotonic \cite{boyd2004}, the fact that informational and quantum entropy productions are convex \cite{cover2006,nielsen2002} implies the idea that their rates are always positive \cite{landi2021}.

Some years ago, we have shown that, for finite-time and weak isothermal driving processes with non-monotonic protocols, the entropy production rate will not be always greater than zero for all times of the process \cite{naze2020,deffner2021}. Inevitably, critics were raised, since the community proclaims a positive entropy production rate as a law of nature \cite{landi2021}. In this work, we try to offer another answer to those critics than the simple difference between definitions of entropy productions \cite{deffner2022}. Indeed, the property of monotonicity persists in our entropy production, but not in the usual sense: we must observe it as a functional in the control protocol and not as a function in time. We show from this property that the Second Law of Thermodynamics is always satisfied, like many other monotonic entropy productions in the literature \cite{callen1998,seifert2005,cover2006,nielsen2002}. We prove the monotonicity by showing that our entropy production is a convex functional \cite{allaire2007}.

At the end, we prove by the convexity of the entropy production the existence of its global optimal protocols \cite{allaire2007}. Therefore, any method that uses the functional of entropy production to minimize it will return such functions. In particular, the analytical method used in Ref. \cite{naze2022} must be in that case. We exemplify that in the same problem of Ref.~\cite{naze2022}, by using a global optimization technique called genetic programming \cite{weise2009}, where the cost function -- our entropy production -- will be minimized by routines of evolutionary selecting processes \cite{koza1992,duriez2016}.  

\section{Linear response theory}
\label{sec:lrt}

We start defining our framework and notations to develop the main concepts to be used in this work.

\subsection{Entropy production}

Consider a classical system of interest, initially in equilibrium with a heat bath of temperature $\beta\equiv {(k_B T)}^{-1}$, where $k_B$ is Boltzmann's constant. The system of interest has a Hamiltonian $\mc{H}(\mb{z}(\mb{z_0},t)),\lambda(t))$, where $\mb{z}(\mb{z_0},t)$ is a point in the phase space evolved from the initial point $\mb{z_0}$ until time $t$, with $\lambda(t)$ being a time-dependent external parameter. Considering that the heat bath does not depend explicitly on $\lambda(t)$, during a switching time $\tau$, the external parameter is changed from $\lambda_0$ to $\lambda_0+\delta\lambda$. The average work performed on the system during this interval of time is \cite{jarzynski1997}
\be
W \equiv \int_0^\tau \la\pd_{\lambda}\mc{H}(t)\ra\dot{\lambda}(t)dt,
\label{eq:work}
\ee
where $\partial_\lambda$ is the partial derivative with respect to $\lambda$, and the superscripted dot is the total time derivative. The generalized force $\la\pd_{\lambda}\mc{H}\ra$ is calculated using the averaging $\langle\cdot\rangle$ over a non-equilibrium probabilistic distribution of the whole system. Consider also that the external parameter can be expressed as
\be
\lambda(t) = \lambda_0+g(t)\delta\lambda,
\ee
where to satisfy the initial conditions of the external parameter, the protocol $g(t)$ must satisfy the following boundary conditions
\be
g(0)=0,\quad g(\tau)=1. 
\label{eq:bc}
\ee
We consider as well that $g(t)\equiv g(t/\tau)$, which means that the intervals of time are measured according to the switching time unit.

Linear-response theory aims to express average quantities until the first order of some perturbation parameter considering how this perturbation affects the observable to be averaged and the non-equilibrium probabilistic distribution \cite{kubo2012}. In our case, we consider that the parameter does not considerably change during the process, $|g(t)\delta\lambda/\lambda_0|\ll 1$, for all $t\in[0,\tau]$. In that manner, using the framework of linear-response theory, the generalized force can be approximated until the first-order as
\begin{equation}
\begin{split}
\la\pd_{\lambda}\mc{H}(t)\ra =&\, \la\pd_{\lambda}\mc{H}\ra_0+\delta\lambda\la\pd_{\lambda\lambda}^2\mc{H}\ra_0 g(t)\\
&-\delta\lambda\int_0^t \phi_0(t-t')g(t')dt',
\label{eq:genforce-resp}
\end{split}
\end{equation}
where $\la\cdot\ra_0$ is the average over the initial canonical ensemble. The quantity $\phi_0(t)$ is the so-called response function \cite{kubo2012}, which can be conveniently expressed as the derivative of the relaxation function $\Psi_0(t)$ \cite{kubo2012}
\be
\phi_0(t) = -\frac{d \Psi_0}{dt}.
\ee 
In our particular case, the relaxation function is calculated as
\be
\Psi_0	(t) = \beta\la\pd_\lambda\mc{H}(0)\overline{\pd_\lambda\mc{H}}(t)\ra_0-\mc{C},
\ee 
where $\overline{\cdot}$ is the stochastic average over the stochastic variables of the system of interest \cite{tome2015}, and the constant $\mc{C}$ is calculated to vanish the relaxation function for long times \cite{kubo2012}. The generalized force, written in terms of the relaxation function, can be expressed as
\begin{equation}
\begin{split}
\la\pd_{\lambda}\mc{H}(t)\ra =&\, \la\pd_{\lambda}\mc{H}\ra_0-\delta\lambda\widetilde{\Psi}_0 g(t)\\
&+\delta\lambda\int_0^t \Psi_0(t-t')\dot{g}(t')dt',
\label{eq:genforce-relax}
\end{split}
\end{equation}
where $\widetilde{\Psi}_0(t)\equiv \Psi_0(0)-\la\pd_{\lambda\lambda}^2\mc{H}\ra_0$. Finally, combining Eqs. (\ref{eq:work}) and (\ref{eq:genforce-relax}), the average work performed at the first-order approximation of the generalized force is
\begin{equation}
\begin{split}
W = &\, \delta\lambda\la\pd_{\lambda}\mc{H}\ra_0-\frac{\delta\lambda^2}{2}\widetilde{\Psi}_0\\
&+\delta\lambda^2 \int_0^\tau\int_0^t \Psi_0(t-t')\dot{g}(t')\dot{g}(t)dt'dt.
\label{eq:work2}
\end{split}
\end{equation}

We observe that the double integral on Eq. (\ref{eq:work2}) vanishes for long switching times \cite{naze2020}. Therefore the other terms are part of the contribution of the difference of free energy since this quantity is exactly the average work performed for quasistatic processes in isothermal drivings. Thus, we can split the average work in the difference of free energy $\Delta F$ and irreversible work $W_{\rm irr}$
\be
\Delta F = \delta\lambda\la\pd_{\lambda}\mc{H}\ra_0-\frac{\delta\lambda^2}{2}\widetilde{\Psi}_0,
\ee  
\begin{equation}
\begin{split}
W_{\text{irr}} = \delta\lambda^2 \int_0^\tau\int_0^t \Psi_0(t-t')\dot{g}(t')\dot{g}(t)dt'dt.
\label{eq:wirrder0}
\end{split}
\end{equation}
In particular, the irreversible work can be rewritten using the symmetric property of the relaxation function, that is, $\Psi_0(t)=\Psi_0(-t)$,
\begin{equation}
\begin{split}
W_{\text{irr}} = \frac{\delta\lambda^2}{2} \int_0^\tau\int_0^\tau \Psi_0(t-t')\dot{g}(t')\dot{g}(t)dt'dt.
\label{eq:wirrLR}
\end{split}
\end{equation}
This irreversible work corresponds to the part of the entropy which is internally raised along the driving since the relaxation function must be a positive kernel \cite{naze2020}. Therefore, the entropy production of the system, given by $S_i=W_{\rm irr}/T$, will be 
\begin{equation}
\begin{split}
S_i = \frac{\delta\lambda^2}{2} \int_0^\tau\int_0^\tau \Psi_0(t-t')\dot{g}(t')\dot{g}(t)dt'dt,
\label{eq:entropyprod}
\end{split}
\end{equation}
where we assume without loss of generality that the temperature of the heat bath is $T=1$. Also, the entropy production rate will be
\begin{equation}
\begin{split}
\dot{S}_i = \frac{\delta\lambda^2}{2} \dot{g}(t)\int_0^\tau \Psi_0(t-t')\dot{g}(t')dt',
\label{eq:dentropyprod}
\end{split}
\end{equation}
which can be positive or negative, depending on the characteristics of the system and process \cite{naze2020}. Another approach to give a physical interpretation of Eq.~\eqref{eq:dentropyprod} is observing it as
\begin{equation}
\begin{split}
\dot{S}_i = \frac{\delta\lambda^2}{2} \left(\int_0^\tau \delta(t-t')\dot{g}(t')dt'\right)\left(\int_0^\tau \Psi_0(t-t')\dot{g}(t')dt'\right),
\label{eq:entropyprod2}
\end{split}
\end{equation}
where the first and second factors are the instantaneous and delayed response of the system in respect to the $\dot{g}$ \cite{naze2020}. Therefore, the entropy production rate can be seen as an instantaneous response modulated appropriately by the delayed response.

\subsection{Diagram of non-equilibrium regions}

\begin{figure}
    \includegraphics[scale=0.4]{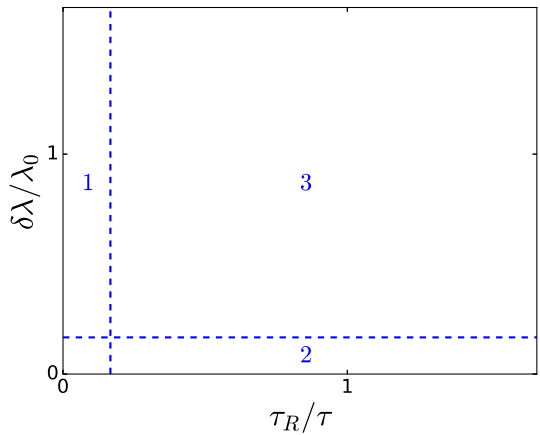}
    \caption{(Color online) Diagram of non-equilibrium regions. Region 1: slowly-varying processes, Region 2: finite-time but weak processes, and Region 3: arbitrarily far-from-equilibrium processes. Linear-response theorem can describe regions 1 and 2. In this work, we are going to focus on region 2 only.}
\label{fig:diagram}
\end{figure}

We establish the regimes where linear-response theory can describe thermodynamic driving processes. Those regimes are determined by the relative strength of the driving with respect to the initial value of the protocol, $\delta\lambda/\lambda_0$, and by the ratio between the relaxation time of the system for the rate by which the process occurs, $\tau_R/\tau$. See Fig.~\ref{fig:diagram} for a diagram depicting the regimes. In region 1, the so-called slowly varying processes, the ratio $\delta\lambda/\lambda_0$ is arbitrary, while $\tau_R/\tau\ll 1$. By contrast, in region 2, the so-called finite-time and weak processes, the ratio $\delta\lambda/\lambda_0\ll 1$, while $\tau_R/\tau$ is arbitrary. In region 3, the so-called arbitrarily far-from-equilibrium processes, both ratios are arbitrary. Linear-response theory is only able to describe regions 1 and 2. In this work, we are going to focus on region 2 only.

\subsection{Overdamped Brownian motion}

To be presented in Sec.~\ref{sec:gmgp}, we describe now the examples of overdamped Brownian motions subjected to time-dependent harmonic traps \cite{seifert2007,naze2022}. Consider a particle of mass $m=1$ and position $x(t)$, subjected to a heat bath and time-dependent harmonic potentials $V(x(t),\lambda(t))$, where $\lambda(t)$ is the external control parameter. Its dynamics are governed by the Langevin equation
\be
m\ddot{x}(t)+\gamma\dot{x}+\pd_t V(x(t),\lambda(t))=\eta(t),
\ee
where $\gamma$ is the friction constant and $\eta(t)$ is a Gaussian white noise, which obeys
\be
\langle\eta(t)\rangle=0,\quad \langle\eta(t)\eta(t')\rangle \propto \delta(t-t').
\ee
We say that the system is subjected to a moving laser trap when
\be
V(x(t),\lambda(t))=\frac{\omega_0^2}{2}{(x(t)-\lambda(t))}^2,
\ee
where $\omega_0$ is the natural frequency of the system. Also, we say that the system is subjected to a stiffening trap when
\be
V(x(t),\lambda(t))=\frac{\lambda(t)}{2}x(t)^2.
\ee
In particular, we treat the regime of overdamped Brownian motion, where we consider the limits $\gamma\rightarrow\infty$ and $\omega_0^2/\gamma<\infty$. In this case, the acceleration term is ignored in comparison to the other terms of the dynamics.

\subsection{Optimization of the entropy production}

Consider the entropy production rewritten in terms of the protocol $g(t)$ instead of its derivative
\be
\begin{aligned}
    S_i =& \delta\lambda^2\frac{\Psi(0)}{2}+\delta\lambda^2\int_0^\tau \dot{\Psi}_0(\tau-t)g(t)dt\\&-\frac{\delta\lambda^2}{2}\int_0^\tau\int_0^\tau \ddot{\Psi}_0(t-t')g(t)g(t')dt dt'.
\end{aligned}
\label{eq:entropyprodg}
\ee
Using the calculus of variations \cite{gelfand2000,kirk2004}, we can derive the Euler-Lagrange equation that gives the optimal protocol of the system to minimize, in principle, the entropy production locally
\be
\int_0^\tau \ddot{\Psi}_0(t-t')g^*(t')dt' = \dot{\Psi}_0(\tau-t).
\label{eq:eleq}
\ee
For applications of the method, see Ref.~\cite{naze2022}.

At this point, to work with the ideas of global minimum and monotonicity of the entropy production, we must look at it as a convex functional in the derivative of the protocol. In particular, this property is proved by showing that the functional is twice Gâteaux differentiable.

\section{G\^ateaux differentiability}
\label{sec:epgdf}

In the following sections, the demonstrations of the main results are omitted. We recommend therefore the book \cite{allaire2007} for a more detailed discussion.

\subsection{G\^ateaux derivative}

Consider an inner product space of functions. The G\^ateaux differentiability is a generalization of the traditional directional derivative applied to functionals defined in that space. In this manner, the functional $J$ is G\^ateaux differentiable at $u$, along the direction $v$, if two things happen: the derivative
\be
    J'[u;v]:=\frac{d J[u+\epsilon v]}{d\epsilon}\Bigg|_{\epsilon=0}
\label{eq:gateaux}
\ee
exists for all $v$ and
\be
    \frac{d J[u+\epsilon v]}{d\epsilon}\Bigg|_{\epsilon=0} = (J'[u],v),
\ee 
where $(\cdot,\cdot)$ is the inner product of that space. The quantity $J'[u]$ is called the G\^ateaux derivative of $J$ in $u$.

Extensions of that concept to higher derivatives exist as well. For example, we say that $J$ is twice G\^ateaux differentiable at $u$ in the direction $v$ and $w$ if the limit
\be
    J''[u;v,w]:=\frac{d J'[u+\epsilon v;w]}{d\epsilon}\Bigg|_{\epsilon=0}
\label{eq:gateaux2}
\ee
exists for all $v$ and $w$, and
\be
    \frac{d J'[u+\epsilon v;w]}{d\epsilon}\Bigg|_{\epsilon=0} = (J''[u]v,w),
\ee
being $J''[u]$ called the second G\^ateaux derivative of $J$ in $u$.

\subsection{Convexity}

In what follows, we present a criterion that connects convexity with twice G\^ateaux differentiability. Starting from the beginning, a functional $J(u)$ is convex if
\be
    J[(1-\theta)u+\theta v] \le (1-\theta)J[u]+\theta J[v],  
\ee
for all $\theta\in[0,1]$, $u$ and $v$. 

In the case where the functional is twice G\^ateaux differentiable, it is convex if, and only if,
\be
(J''[u]w,w)\ge 0,
\ee
for all $u$ and $w$ \cite{allaire2007}. In the following, we briefly present, from convexity, results involving monotonicity and global minimization.

\subsection{Monotonicity of G\^ateaux derivative}

A functional $J(u)$ is convex if, and only if, its G\^ateaux derivative is monotonic \cite{allaire2007}
\be
(J'[u]-J'[v],u-v)\ge 0.
\ee
This property will show that the entropy production preserves the idea of monotonicity \cite{naze2020}, but in the sense of a functional depending on the derivative of the protocol.

\subsection{Global optimization}

If a functional is convex, every local minimum will be a global minimum \cite{allaire2007}. Suppose for instance that $u$ is a local minimum. If it is not a global minimum, it exists a $v$ such that $J[v]<J[u]$. By convexity, it holds
\be
J[(1-\theta)u+\theta v]<J[u],
\ee
for sufficiently small $\theta$, $J(u)$ is not a local minimum, which contradicts the hypothesis. 

We apply those concepts presented at the functional of entropy production derived from linear-response theory for weak drivings.

\section{Monotonicity of entropy production}

We are going to show that the functional of the entropy production $S_i[\dot{g}(t)]$ is monotonic in $\dot{g}(t)$ in the sense of functionals. We use the idea of convexity of $S_i$. First, we observe that Eq.~\eqref{eq:entropyprod} defines a natural inner product of the space of functions
\be
(\dot{g}(t),\dot{h}(t)) = \int_0^\tau\left(\int_0^\tau \Psi_0(t-t')\dot{g}(t')dt'\right)\dot{h}(t)dt,
\ee
where $\Psi(t)$ is the relaxation function. In Ref.~\cite{naze2020}, we have shown that such relaxation function must be a positive kernel, such that the inner product becomes well-defined. The first G\^ateaux derivative of $S_i$ with respect to $\dot{g}(t)$ is
\be
S'_i[\dot{g}(t)] = \int_0^\tau \Psi_0(t-t')\dot{g}(t')dt',
\ee
and the second one
\be
S_i''[\dot{g}(t)] = 1.
\label{eq:twicediff}
\ee
In that manner, since Eq.~\eqref{eq:twicediff} is a positive number, $S_i$ is a convex functional. Therefore, $S_i[\dot{g}(t)]$ must be monotonic
\be
(S_i'[\dot{g}_1(t)]-S_i'[\dot{g}_2(t)],\dot{g}_1(t)-\dot{g}_2(t) )\ge 0,
\ee
which means
\be
\int_0^\tau\int_0^\tau\Psi_0(t-t')(\dot{g}_1(t)-\dot{g}_2(t))(\dot{g}_1(t')-\dot{g}_2(t'))dtdt'\ge 0,
\label{eq:mono}
\ee
for all $\dot{g}_1(t)$ and $\dot{g}_2(t)$. Equation \eqref{eq:mono} implies, by Bochner's theorem \cite{naze2020}, that the relaxation function must have its Fourier transform positive, as it was supposed at the beginning in the definition of the inner product. Indeed, the monotonicity guarantees such property without taking it for granted. In this manner, the physical consequence of the monotonicity of the G\^ateaux derivative is the Second Law of Thermodynamics. This is in complete agreement with many other types of entropy productions whose monotonicities imply the satisfaction of the Second Law of Thermodynamics
\cite{callen1998,seifert2005,cover2006, nielsen2002,landi2021}.
\begin{figure}
    \includegraphics[scale=0.50]{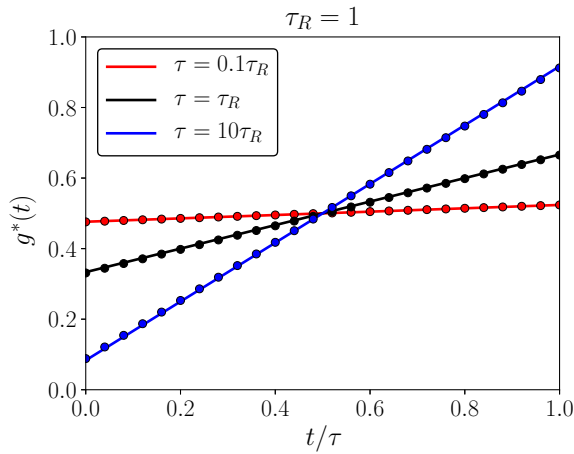}
    \caption{(Color online) Comparison between genetic programming (circles) and analytical (solid lines) results in different switching times. The matching between them exemplifies that the analytical optimal protocols are global solutions. Our criterion of convergence was achieved when graphically the protocol does not change significantly for 30 generations.}
\label{fig:over}
\end{figure}

What exactly does it add to our knowledge this monotonicity of the entropy production besides the satisfaction of the Second Law of Thermodynamics? Since Jarzynski \cite{jarzynski1997} recovers this result from its equality for processes of any strength it seems unnecessary to reprove this fact. But this is not the case: the result informs the roots about how the forces correlate at two different times along the process by which the system passes. Indeed, the positive Fourier transform property, assumed before {\it ad hoc} to the relaxation function \cite{naze2020}, is a direct consequence of the intrinsic convexity of the entropy production.

\section{Global minimum for entropy production}

Another important consequence of $S_i$ being convex is that any local minimum will be the global minimum \cite{allaire2007}. In this manner, the optimal protocols calculated in Ref.~\cite{naze2022} by solving Eq.~\eqref{eq:eleq} are indeed global minimum. We verify that by comparing the analytical results with a global optimization technique called genetic programming \cite{weise2009,duriez2016,koza1992}.

\subsection{Global minimum with genetic programming}
\label{sec:gmgp}

Consider the global optimization method of genetic programming \cite{weise2009,koza1992,duriez2016}. It consists basically of finding the optimal protocol evaluating, among members of a family of functions, the minimal cost functional using evolutionary selecting routines along generations. Using the MATLAB package MCL2 \cite{duriez2016}, we develop a code to find the optimal protocol to minimize the entropy production functional of an overdamped Brownian motion subjected to time-dependent harmonic traps \cite{seifert2007,naze2022}. In both cases of moving laser and stiffening traps, the relaxation function is
\be
\Psi(t) \propto e^{-|t|/\tau_R},
\ee
where $\tau_R$ is the characteristic relaxation timescale of the problem. The cost function of genetic programming will be Eq. \eqref{eq:entropyprod}. We expect that the code converges to the global optimal protocol unless it presents an error in their algorithm or a bad choice of parameters \cite{weise2009}. In this manner, the simulations were repeated several times with different but reasonable initial conditions and parameters \cite{duriez2016}. Except for the final solution, no momentary stop in a particular protocol was found within our criterion of convergence in any simulation. This highly suggests the nonexistence of local minima different than that of the global one. Our criterion of convergence was achieved when graphically the protocol does not change significantly for 30 generations.

On the other hand, the analytical optimal protocols of this problem are given by \cite{seifert2007,naze2022}
\be
g^{*}(t) = \frac{t+\tau}{\tau+2\tau_R}.
\ee

The comparison between genetic programming and analytical results is depicted in Fig.~\ref{fig:over}. The matching between both results illustrates that the analytical optimal protocol is indeed a global minimum, as predicted by its property of convexity. 

We remark that we have proven the convexity for $\dot{g}(t)$, and not for $g(t)$. However, the transformation of Eq.~\eqref{eq:entropyprod} to Eq.~\eqref{eq:entropyprodg} will not change the property of convexity. Indeed, calculating the second G\^ateaux derivative of Eq.~\eqref{eq:entropyprodg} it will be positive, since $\ddot{\Psi}_0$ is a negative kernel \cite{naze2020}. It remains the question if the optimal derivative of the protocol will be the derivative of the optimal one. Indeed, expressing the Euler-Lagrange equation of Eq.~\eqref{eq:entropyprod} 
\be
\int_0^\tau\int_0^\tau \Psi(t-t')\dot{g}(t)\dot{h}(t)dt'dt=0
\ee
in terms of a combination of $g(t)$ and $h(t)$, we will arrive in the same Euler-Lagrange equation of Eq.~\eqref{eq:entropyprodg}. Therefore, the solutions are equal.

Observe also that our method is not only restricted to standard white noise overdamped Brownian motion, or even isothermal process: the functional \eqref{eq:entropyprod} always guarantees convexity. Inertial particles \cite{caprini2021}, Ornstein-Uhlenbeck processes \cite{caprini2019}, or thermally isolated systems \cite{acconcia2015,acconcia2015b,bonanca2016} may be good examples to verify global optimization. At this point, however, we do not possess a general solution of the Euler-Lagrange equation for their relaxation functions, making a comparison with genetic programming results unfeasible.

Finally, the existence of a global minimum is guaranteed since for relaxation functions -- which are symmetric kernels -- there is a solution for the Euler-Lagrange equation, which can be implemented from eigenvalues and eigenvectors problem of the integral equation \cite{manzhirov2008}. The uniqueness however is not always true, since for degenerate relaxation functions there is an infinite number of solutions for the Euler-Lagrange equation \cite{manzhirov2008}. The relaxation function of cosine is a typical example \cite{acconcia2015b}.

\section{Final remarks}
\label{sec:final} 

We proved that, for finite-time and weak isothermal driving processes, entropy production is a convex functional in the derivative of the protocol. Therefore, this quantity presents a global optimal protocol, and it is monotonic. We verified the first consequence by comparing the results of the analytical method presented in Ref.~\cite{naze2022} with those of genetic programming presented in Sec.~\ref{sec:gmgp}. In the second consequence, the entropy production, by contrast with our previous works \cite{naze2020,deffner2021}, is monotonic when seen as a functional in an inner product space of functions. This property implies the satisfaction of the Second Law of Thermodynamics, agreeing therefore with other monotonic entropy productions proposed in the literature in other contexts. Other optimization methods using the entropy production functional \eqref{eq:entropyprod}, like the minimization of a finite quadratic form with Lagrange multipliers \cite{bonanca2018,kamizaki2022,soriani2022}, present as well the same property of having a global optimal protocol.

\begin{acknowledgements}
The author acknowledges Marcus V. S. Bonan\c{c}a and Sebastian Deffner for fruitful discussions.
\end{acknowledgements}

\bibliography{bibliography}

\end{document}